\documentclass[reprint, amssymb, amsmath, aps, superscriptaddress, showpacs, prb]{revtex4-1}
\usepackage{graphicx}
\usepackage{hyperref}

\usepackage{graphicx}
\usepackage{dcolumn}
\usepackage{amsmath}
\usepackage{color}
\usepackage{ulem}
\usepackage{multirow}

\begin{document}

\title{Two-gap superconductivity in Mo$_{8}$Ga$_{41}$ and its evolution upon the V substitution}
\author{V.Yu. \surname{Verchenko}}
\affiliation{Department of Chemistry, Lomonosov Moscow State University, 119991 Moscow, Russia}
\email{verchenko@inorg.chem.msu.ru}
\affiliation{National Institute of Chemical Physics and Biophysics, 12618 Tallinn, Estonia}

\author{R. \surname{Khasanov}}
\affiliation{Laboratory for Muon Spin Spectroscopy, Paul Scherrer Institute, CH-5232 Villigen PSI, Switzerland}
\email{rustem.khasanov@psi.ch}

\author{Z. \surname{Guguchia}}
\affiliation{Laboratory for Muon Spin Spectroscopy, Paul Scherrer Institute, CH-5232 Villigen PSI, Switzerland}

\author{A.A. \surname{Tsirlin}}
\affiliation{Experimental Physics VI, Center for Electronic Correlations and Magnetism, Institute of Physics, University of Augsburg, 86135 Augsburg, Germany}

\author{A.V. \surname{Shevelkov}}
\affiliation{Department of Chemistry, Lomonosov Moscow State University, 119991 Moscow, Russia}

\begin{abstract}
Zero-field and transverse-field muon spin rotation/relaxation ($\mu$SR) experiments were undertaken in order to elucidate microscopic properties of a strongly-coupled superconductor Mo$_{8}$Ga$_{41}$ with $T_{\text{c}}=9.8$\,K. The upper critical field extracted from the transverse-field $\mu$SR data exhibits significant reduction with respect to the data from thermodynamic measurements indicating the coexistence of two independent length scales in the superconducting state. Accordingly, the temperature-dependent magnetic penetration depth of Mo$_{8}$Ga$_{41}$ is described using the model, in which two \textit{s}-wave superconducting gaps are assumed. The V for Mo substitution in the parent compound leads to the complete suppression of one superconducting gap, and Mo$_{7}$VGa$_{41}$ is well described within the single \textit{s}-wave gap scenario. The reduction in the superfluid density and the evolution of the low-temperature resistivity upon the V substitution indicate the emergence of a competing state in Mo$_{7}$VGa$_{41}$ that may be responsible for the closure of one of the superconducting gaps.
\end{abstract}

\pacs{74.25.Bt, 74.70.Ad, 76.75.+i}

\maketitle

\section{Introduction}

Design of new superconducting materials following empirical rules based on chemical and structural considerations or rigorous analysis of the electronic structure is a challenging task. Recently, systematic work was performed on superconducting intermetallics with endohedral gallium clusters TGa$_{\text{n}}$ (T is a transition metal)\cite{cava}. In this context, Mo$_{8}$Ga$_{41}$\cite{m} (the superconducting transition temperature $T_{\text{c}}\simeq9.7$\,K), Mo$_{6}$Ga$_{31}$\cite{poole} ($T_{\text{c}}\simeq8$\,K), ReGa$_{5}$\cite{cava} ($T_{\text{c}}\simeq2.3$\,K), Rh$_{2}$Ga$_{9}$\cite{t2ga9} ($T_{\text{c}}\simeq1.9$\,K) and Ir$_{2}$Ga$_{9}$\cite{t2ga9} ($T_{\text{c}}\simeq2.2$\,K) superconductors as well as non-superconducting V$_{8}$Ga$_{41}$\cite{v8ga41} and PdGa$_{5}$\cite{pdga5} are considered. By analyzing their electronic structures, Xie \textit{et al.}\cite{cava} established the interrelations between the critical temperature $T_{\text{c}}$ and valence electron count, thus placing these structurally different intermetallic compounds into an individual class of superconductors. 

\begin{figure}
\includegraphics{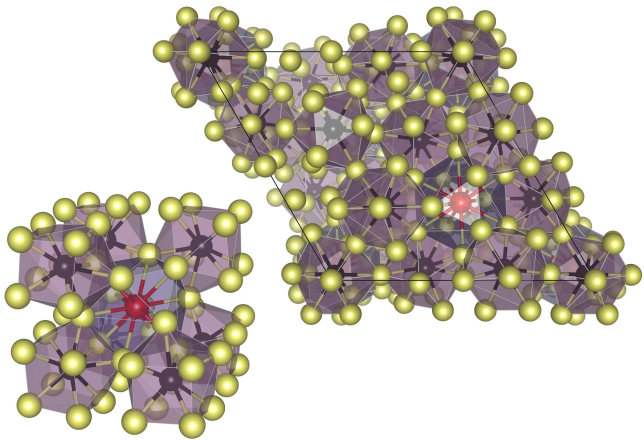}
\caption{\label{f1}View of the Mo$_{8}$Ga$_{41}$ crystal structure: (left) eight MoGa$_{10}$ polyhedra surrounding the GaGa$_{12}$ cuboctahedron (one polyhedron is not shown for clarity); (right) polyhedral representation of the unit cell. Mo atoms are shown in black color, Ga atoms in yellow color, and the unique Ga atom in the center of cuboctahedron -- in red color.}
\end{figure}

Mo$_{8}$Ga$_{41}$ was firstly synthesized by Yvon \textit{et al.}\cite{yvon} during their systematic investigation of Ga-rich intermetallic compounds with unusual crystal structures. View of the Mo$_{8}$Ga$_{41}$ crystal structure is shown in Fig.~\ref{f1}. The crystal structure is built by MoGa$_{10}$ polyhedra, which are condensed on the triangular faces of GaGa$_{12}$ cuboctahedron. A detailed description of the crystal structure can be found elsewhere\cite{our}. From the first sight, Mo$_{8}$Ga$_{41}$ follows all of the Matthias' empirical rules for superconductors: it contains a 4\textit{d} transition metal (Mo), the compound may possess high density of  electronic states, and its complex crystal structure is closely related to the cubic symmetry. Despite the fact that nowadays these rules seem to be obsolete, Yvon and co-authors\cite{m} found that Mo$_{8}$Ga$_{41}$ superconducts below $T_{\text{c}}=9.7$\,K with the upper critical field of $\mu_{0}H_{\text{c2}}=8.6$\,T. Results obtained by Yvon \textit{et al.} were further pointed out on the unusual behavior of Mo$_{8}$Ga$_{41}$ in its superconducting state. Indeed, values of $T_{\text{c}}$ and $dH_{\text{c2}}/dT$ lead to the Werthamer-Helfand-Honenberg (WHH) prediction of the upper critical field $\mu_{0}H_{\text{c2}}\simeq6.7$\,T, which is significantly smaller than the experimental value\cite{m}. Together with a relatively high critical temperature, this may indicate strong electron-phonon coupling in the superconducting regime.

In our recent study\cite{our}, we carried out comprehensive investigation of  thermodynamic and transport properties of Mo$_{8}$Ga$_{41}$ in both, the normal and superconducting, states. Magnetization, heat capacity, and resistivity measurements confirmed bulk superconductivity at $T_{\text{c}}=9.8$\,K in zero magnetic field. The superconducting transition was found to be characterized by a large value of the normalized specific heat jump $\Delta{}c_{\text{p}}/\gamma_{\text{N}}T_{\text{c}}=2.84$, which is twice as high as the weak-coupling BCS prediction. At the same time, the electronic contribution to the specific heat below $T_{\text{c}}$ obeyed a power-law rather than exponential decay behavior as is expected within the framework of BCS theory. Such disagreements have motivated us for further investigation of microscopic properties of Mo$_{8}$Ga$_{41}$.

Here, we present the results of muon spin rotation/relaxation ($\mu$SR) experiments on bulk Mo$_{8}$Ga$_{41}$ and its V-substituted derivative, Mo$_{7}$VGa$_{41}$. Transverse-field $\mu$SR (TF-$\mu$SR) experiments were undertaken in order to elucidate superconducting-state properties of both materials. The temperature-dependent magnetic penetration depth $\lambda$ and the upper critical field $\mu_{0}H_{\text{c2}}$ were extracted from the TF-$\mu$SR data and were used to investigate the order parameter in Mo$_{8}$Ga$_{41}$ and its evolution upon the V for Mo substitution.

\section{Experimental details}
Mo$_{8}$Ga$_{41}$ and Mo$_7$VGa$_{41}$ specimens were synthesized as described earlier\cite{our} in the form of small submillimeter-size crystals. Elemental composition of the obtained crystals was investigated using a JSM~JEOL~6490-LV scanning electron microscope operated at 30\,kV and equipped with an energy-dispersive x-ray detection system INCA~x-Sight. For quantitative analysis, molybdenum and vanadium elements, and gallium phosphide provided by MAC Analytical Standards were used as external standards. According to the EDX spectroscopy results, crystals of Mo$_{8}$Ga$_{41}$ and Mo$_{7}$VGa$_{41}$ contain 16.3(5)\,at.\,$\%$ Mo, 83.7(8)\,at.\,$\%$ Ga, and 14.4(5)\,at.\,$\%$ Mo, 2.0(2)\,at.\,$\%$ V, and 83.6(9)\,at.\,$\%$ Ga, respectively, leading to the Mo$_{8.0(2)}$Ga$_{41.0(4)}$ and Mo$_{7.1(2)}$V$_{1.0(1)}$Ga$_{41.0(4)}$ formula units. In the case of Mo$_{7}$VGa$_{41}$\cite{sm}, elemental mapping was used to check the distribution of Mo and V species across the surface, which was found to be uniform confirming local homogeneity of crystals. For measurements of thermodynamic properties, several crystals of Mo$_{7}$VGa$_{41}$ were glued together and measured as a polycrystalline sample. Magnetization curves were registered in magnetic fields between 0\,T and 14\,T at temperatures between 1.8\,K and 10\,K using the VSM setup of a Physical Property Measurement System (PPMS, Quantum Design). Heat capacity was measured in magnetic fields up to 12\,T by the relaxation method using the Heat Capacity option of PPMS. Further, crystals of Mo$_{8}$Ga$_{41}$ and Mo$_{7}$VGa$_{41}$ were crushed by grinding in an agate mortar and analyzed by powder x-ray diffraction technique. The experiments were performed on a Huber Guinier Camera G670 [Image plate detector, Cu x-ray source, Ge (111) monochromator, $\lambda=1.540598$\,\r A]. PXRD patterns of Mo$_{8}$Ga$_{41}$ and Mo$_{7}$VGa$_{41}$ agree with the V$_{8}$Ga$_{41}$ structure type and show no impurity phases in the specimens. Finally, the obtained powders of Mo$_{8}$Ga$_{41}$ and Mo$_{7}$VGa$_{41}$ were pressed into cylindrical pellets with the diameter of 6\,mm and the height of 1\,mm at room temperature at the external pressure of 100\,bar. These pellets were used for the $\mu$SR experiments.

$\mu$SR measurements were performed at the Paul Scherrer Institute (PSI), Villigen, Switzerland. Zero-field (ZF) and transverse-field (TF) $\mu$SR experiments were carried out on the GPS and DOLLY spectrometers located at the $\pi$M3 and $\pi$E1 beamlines, respectively. Experiments were performed in the temperature range from 1.6 to 15~K in GPS and from 0.29 to 15~K in DOLLY instruments. ZF-$\mu$SR experiments were performed in zero applied field. In TF-$\mu$SR experiments the sample was field-cooled from above $T_c$ in series of fields ranging from 30~mT to 0.59~T.

\section{Results and Discussion}
\subsection{ Mo$_{8}$Ga$_{41}$} 
\subsubsection{Zero-field $\mu$SR experiments}

\begin{figure}
\includegraphics{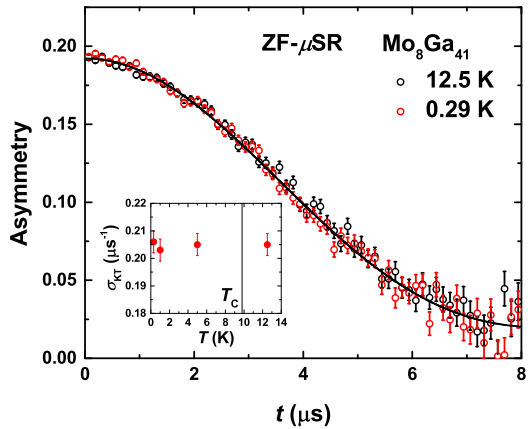}
\caption{\label{f2} ZF-$\mu$SR time-spectra of Mo$_{8}$Ga$_{41}$ measured at $T=0.29$\,K and 12.5\,K. Solid line is a fit of the $T=12.5$\,K data according to the Kubo-Toyabe relation (see in the text). The inset shows zero-field muon depolarization rate $\sigma_{\text{KT}}$ as a function of temperature.}
\end{figure}

Figure~\ref{f2} shows ZF-$\mu$SR time-spectra of Mo$_{8}$Ga$_{41}$ measured at $T=0.29$\,K and 12.5\,K. The obtained spectra below and above $T_{\text{c}}$ reveal no significant difference indicating the absence of internal coherent magnetic fields in Mo$_{8}$Ga$_{41}$ that may appear, for instance, as a result of long-range magnetic ordering. On the contrary, our ZF-$\mu$SR data show that the only nuclear component is present, which can be analyzed by using the static Kubo-Toyabe depolarization function\cite{kt}:

\begin{equation}
A(t)=A_{\text{s}}(0)\left[\frac{1}{3}+\frac{2}{3}\left(1-\sigma_{\text{KT}}^{2}t^{2}\right)\exp\left(-\frac{\sigma_{\text{KT}}^{2}t^{2}}{2}\right)\right]+A_{\text{bgd}}\label{eq1},
\end{equation}

\noindent{}where $A_{\text{s}}(0)$ is the initial asymmetry for muons stopped in the sample, $\sigma_{\text{KT}}$ is the muon depolarization rate, and $A_{\text{bgd}}$ is the background contribution from muons that missed the sample. The muon depolarization rate $\sigma_{\text{KT}}$ does not change with temperature within its standard deviation (see the inset in Fig.~\ref{f2}). This clearly indicates the absence of any spontaneous coherent magnetic fields in the superconducting state of Mo$_{8}$Ga$_{41}$.

\subsubsection{Transverse-field $\mu$SR experiments}

\begin{figure}
\includegraphics{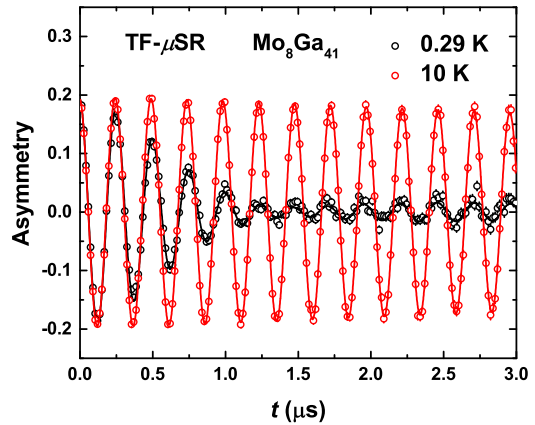}
\caption{\label{f3}TF-$\mu$SR time-spectra of Mo$_{8}$Ga$_{41}$ measured in $\mu_{0}H=30$\,mT at temperatures below (black dots) and above (red dots) the superconducting transition temperature. Solid lines are least-squares fits according to Eq.~\eqref{eq2}.}
\end{figure}

In order to elucidate microscopic properties of the superconducting state in Mo$_{8}$Ga$_{41}$, including the structure of superconducting gap, transverse-field $\mu$SR experiments were carried out in magnetic fields of 30, 70, 200, and 490\,mT. Figure~\ref{f3} shows TF-$\mu$SR time-spectra of Mo$_{8}$Ga$_{41}$ measured at $T=0.29$\,K and 10\,K in $\mu_{0}H=30$\,mT. The spectrum below $T_{\text{c}}$ clearly exhibits faster relaxation with respect to the spectrum at 10\,K due to the formation of the flux line lattice in the superconducting state. The following oscillatory decaying Gaussian function was used to fit the experimental data (the fitting results are shown as solid lines in Fig.~\ref{f3}):

\begin{align}
A^{\text{TF}}(t)=& A^{\text{TF}}_{\text{s}}(0)\times\exp\left(-\frac{\sigma^{2}t^{2}}{2}\right)\times\cos(\gamma_{\mu}\mu_{0}H_{\text{int}}t+\phi)+\notag\\
& A^{\text{TF}}_{\text{bgd}}(0)\times\cos(\gamma_{\mu}\mu_{0}H_{\text{bgd}}t+\phi)\label{eq2}.
\end{align}

\noindent{}Here, $A^{\text{TF}}_{\text{s}}(0)$ and $A^{\text{TF}}_{\text{bgd}}(0)$ are initial asymmetries belonging to the sample and the background contributions, respectively. $\gamma_{\mu}/2\pi=135.5$\,MHz/T is the muon gyromagnetic ratio, $\mu_{0}H_{\text{int}}$ and $\mu_{0}H_{\text{bgd}}$ are the internal and background magnetic fields, respectively, $\phi$ is the initial phase, and $\sigma$ is the Gaussian muon spin relaxation rate. Above $T_{\text{c}}$, the muon spin relaxation rate is referred to the temperature independent nuclear magnetic dipolar contribution $\sigma_{\text{nm}}$, while below $T_{\text{c}}$, the total relaxation rate consists of the nuclear and superconducting contributions, $\sigma=\sqrt{\sigma^{2}_{\text{sc}}+\sigma^{2}_{\text{nm}}}$. By fitting the data at $T=10$\,K, we obtained $\sigma_{\text{nm}}=0.16$\,$\mu{}\text{s}^{-1}$. This value was further used to extract $\sigma_{\text{sc}}(T)$ and $\mu_{0}H_{\text{int}}(T)$ at temperatures below $T_{\text{c}}$ according to Eq.~\ref{eq2}. In the superconducting state, Fourier transforms of the asymmetry spectra are well described by the simple Gaussian approximation what is normally the case for powder samples of anisotropic superconductors. Fourier analysis of the asymmetry spectra shows no indications of disorder of the vortex lattice.

\begin{figure}
\includegraphics{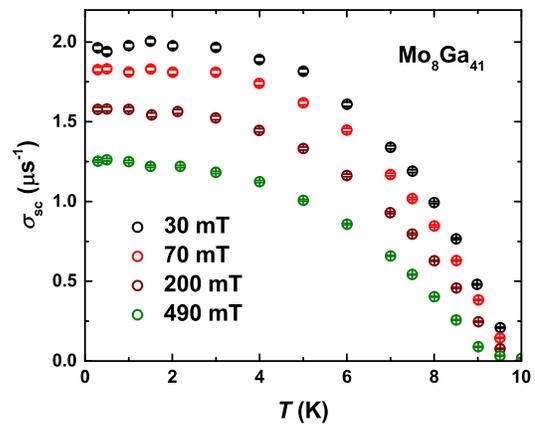}
\caption{\label{f4}Superconducting contribution $\sigma_{\text{sc}}$ to the muon spin relaxation rate of Mo$_{8}$Ga$_{41}$.}
\end{figure}

\begin{figure*}
\includegraphics{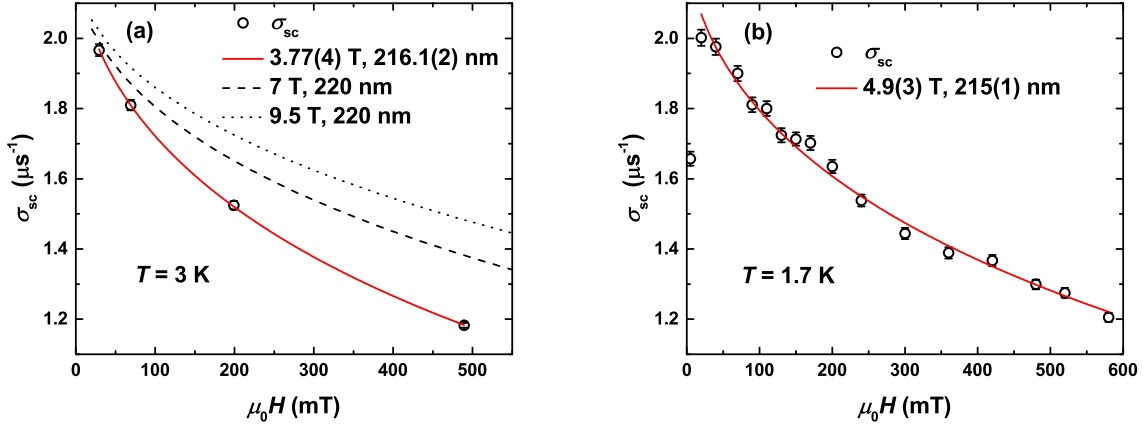}
\caption{\label{f5}(a) Field-dependent superconducting contribution $\sigma_{\text{sc}}$ to the total muon spin relaxation rate of Mo$_{8}$Ga$_{41}$ taken from the TF-$\mu$SR $T$-scans at $T=3$\,K. Solid red line is a fit of the data according to Eq.~\eqref{eq3} with $\mu_{0}H_{\text{c2}}$ and $\lambda$ treated as variable parameters, dashed and dotted lines are the results for $\mu_{0}H_{\text{c2}}=7$\,T and $\mu_{0}H_{\text{c2}}=9.5$\,T, respectively, and $\lambda=220$\,nm. (b) $\sigma_{\text{sc}}$ obtained from the TF-$\mu$SR $\mu_{0}H$-scan at $T=1.7$\,K. The solid red line is a fit of the data according to Eq.~\eqref{eq3}.}
\end{figure*}

From the data taken at each temperature below the transition temperature $T_{\text{c}}$, we reconstructed $\sigma_{\text{sc}}$ as a function of transverse field for $\mu_{0}H=30$, 70, 200 and 490\,mT. The resulting $\sigma_{\text{sc}}(T)$ dependencies for various transverse fields are shown in Fig.~\ref{f4}. The temperature-dependent variation of $\sigma_{\text{sc}}$ exhibits a clear enhancement below $T_{\text{c}}$. Figure~\ref{f4} also implies that $\sigma_{\text{sc}}$ measured at similar temperatures decreases with increasing measuring field [see Fig.~\ref{f5}(a) where $\sigma_{\text{sc}}(\mu_{0}H)$ at $T=3$\,K is depicted].  The more precise scan performed at $T=1.7$\,K shows that before coming down, $\sigma_{\text{sc}}$ first increases by going through a maximum at around 20\,mT [Fig.~\ref{f5}(b)]. Note that this is the typical behavior observed in various type-II superconductors experimentally\cite{cardwell} and as well as predicted theoretically\cite{brandt}.

In terms of Ginzburg-Landau treatment of the vortex state, Brandt\cite{brandt} has shown that $\sigma_{\text{sc}}$ can be calculated as a function of the reduced field $b$ and the Ginzburg-Landau coefficient $\kappa$ for the case of single-gap \textit{s}-wave superconductivity:

\begin{equation}
\sigma_{\text{sc}}\approx0.172\frac{1-b}{\kappa^{2}}\left[1+1.21(1-\sqrt{b})^{3}\right]\label{eq3},
\end{equation}

\noindent{}where $b=\frac{\mu_{0}H}{\mu_{0}H_{\text{c2}}}$, $\mu_{0}H_{\text{c2}}$ is the upper critical field, $\kappa=\lambda/\xi$, $\lambda$ is the magnetic penetration depth, and $\xi$ is the Ginsburg-Landau coherence length. Eq.~\eqref{eq3} is valid for $b$ exceeding $0.25/\kappa^{1.3}$ (where the maximum of $\sigma_{\text{sc}}$ occurs) and   $\kappa\gtrsim5$. For Mo$_{8}$Ga$_{41}$, the value of $\xi=15.8$\,nm was estimated from the upper critical field at zero temperature\cite{our}, thus, Eq.~\eqref{eq3} can be used for $\lambda>79$\,nm.

The $\sigma_{\text{sc}}(\mu_{0}H)$ data obtained from $T$-scans [Fig.~\ref{f5}(a)] as well as from the $\mu_{0}H$-scan [Fig.~\ref{f5}(b)] were fitted using Eq.~\eqref{eq3}. Firstly, the fitting with $\mu_{0}H_{\text{c2}}$ and $\lambda$ treated as variable parameters (solid red lines in Fig.~\ref{f5}) was employed yeilding the values of upper critical field $\mu_{0}H_{\text{c2}}$ that are significantly smaller than that obtained from thermodynamic measurements previously\cite{our}. Indeed, as shown in Fig.~\ref{f5}(a), by fitting the $\sigma_{\text{sc}}(\mu_{0}H)$ data we obtained $\mu_{0}H_{\text{c2}}=3.77(4)$\,T and $\lambda=216.1(2)$\,nm at $T=3$\,K, compared to 6.7\,T from magnetization and heat-capacity measurements\cite{our}. On the other hand, if such a high value of $\mu_{0}H_{\text{c2}}$ is adopted, $\sigma_{\text{sc}}$ has to be significantly enhanced, in poor agreement with the $\mu$SR data. The same situation is observed in the case of $\mu_{0}H$-scan [Fig.~\ref{f5}(b)], where the values of $\mu_{0}H_{\text{c2}}=4.9(3)$\,T and $\lambda=215(1)$\,nm were obtained at $T=1.7$\,K.

\begin{figure}
\includegraphics{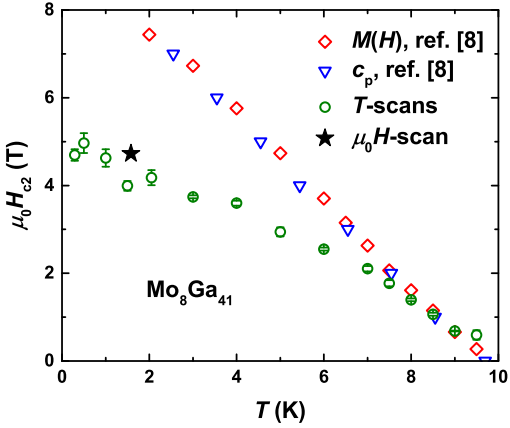}
\caption{\label{f6}Upper critical field $\mu_{0}H_{\text{c2}}$ of Mo$_{8}$Ga$_{41}$ obtained from the $T$-scans and $\mu_{0}H$-scan. Values from the thermodynamic measurements\cite{our} are also shown.}
\end{figure}

Values of the upper critical field of Mo$_{8}$Ga$_{41}$ are summarized in Fig.~\ref{f6}. Obviously, the entire plot of $\mu_{0}H_{\text{c2}}(T)$ calculated from the TF-$\mu$SR data exhibits large reduction in the upper critical field compared to thermodynamic measurements. This situation is a characteristic feature of multigap superconductivity that leads to two or more distinct length scales in the superconducting state\cite{lee}. Therefore, we further studied magnetic penetration depth in order to clarify this multigap behavior.

\subsubsection{Magnetic penetration depth}

\begin{figure}
\includegraphics{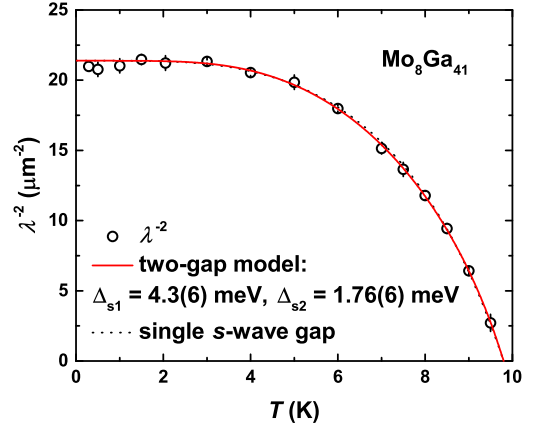}
\caption{\label{f7}Temperature-dependent magnetic penetration depth $\lambda$ of Mo$_{8}$Ga$_{41}$ plotted as $\lambda^{-2}$ \textit{vs.} $T$. Solid line is a least-squares fit according to the two-gap model, the dashed line is a fit according to the single-gap BCS-type model.}
\end{figure}

The temperature-dependent inverse squared magnetic penetration depth $\lambda^{-2}$ of Mo$_{8}$Ga$_{41}$ obtained from the TF-$\mu$SR data using Eq.~\eqref{eq3} is shown in Fig.~\ref{f7}. Upon increasing temperature, $\lambda$ increases monotonically and tends to infinity at $T=T_{\text{c}}$. Concurrently, $\lambda^{-2}$ decreases with increasing temperature and vanishes at the transition temperature. The temperature dependence of $\lambda^{-2}$ is indicative of the order parameter, i.e., absolute value of the gap energy, as well as symmetry of the superconducting gap. Assuming weak-coupling BCS-type \textit{s}-wave superconductivity with a single gap $\Delta$ reveals\cite{tinkham}

\begin{equation}
\frac{\lambda^{-2}(T)}{\lambda^{-2}(0)}=\frac{\Delta(T)}{\Delta(0)}\text{tanh}\left[\frac{\Delta(T)}{2k_{\text{B}}T}\right]\label{eq4},
\end{equation}

\noindent{}in the dirty limit. At non-zero temperatures, the gap $\Delta(T)$ is assumed to follow the function $\Delta(T)=\Delta(0)\,{\rm tanh}[1.82(1.018(T_{\text{c}}/T-1))^{0.51}]$. This function has been found to well represent the temperature dependence at any coupling strength\cite{padamsee}.

The fit using Eq.~\eqref{eq4} is shown in Fig.~\ref{f7} as a dashed line. The single-gap BCS-type model satisfactorily describes the experimental data with $\Delta(0)=1.80(7)$\,meV, $\Delta(0)/k_{B}T_{\text{c}}=2.1$, and $\lambda(0)=192(2)$\,nm. The reduced value of $\chi^{2}=0.66$ has been obtained with $n=13$ degrees of freedom. The results of fitting, from the first sight, hint at the single-gap BCS-type superconductivity in Mo$_{8}$Ga$_{41}$. However, previously we established two non-BCS-type features\cite{our}: (i) the superconducting state is characterized by the strong electron-phonon coupling with $\lambda_{\text{ep}}=0.9$, (ii) the normalized specific heat jump $\Delta{}c_{\text{p}}/\gamma_{\text{N}}T_{\text{c}}=2.84$ is much larger than in the weak-coupling BCS limit. Also, in the current study we found that (iii) the upper critical field from the TF-$\mu$SR data shows the significant reduction with respect to the thermodynamic data. These features suggest that the single-gap BCS-type model should be avoided, and the multigap model should be used instead.

The two-gap model can be introduced using the linear combination of two \textit{s}-wave superconducting gaps:

\begin{align}
\frac{\lambda^{-2}(T)}{\lambda^{-2}(0)}=\omega{}\frac{\Delta_{\text{s1}}(T)}{\Delta_{\text{s1}}(0)}\text{tanh}\left[\frac{\Delta_{\text{s1}}(T)}{2k_{\text{B}}T}\right]+\notag\\
(1-\omega)\frac{\Delta_{\text{s2}}(T)}{\Delta_{\text{s2}}(0)}\text{tanh}\left[\frac{\Delta_{\text{s2}}(T)}{2k_{\text{B}}T}\right]\label{eq5},
\end{align}

\noindent{}where $\omega$ is the linear combination coefficient describing the contribution of each gap, and $\Delta_{\text{s}i}(T)=\Delta_{\text{s}i}(0)\,{\rm tanh}[1.82(1.018(T_{\text{c}}/T-1))^{0.51}]$ is the temperature dependence of each gap. Fit of the $\lambda^{-2}(T)$ data using Eq.~\ref{eq5} is shown as a solid red line in Fig.~\ref{f7} yielding $\Delta_{\text{s1}}(0)=4.3(6)$\,meV, $\Delta_{\text{s2}}(0)=1.76(6)$\,meV, and $\lambda(0)=216.2(4)$\,nm. This fitting is characterized by the reduced value of $\chi^{2}=0.50$ with $n=12$ degrees of freedom for $\omega=0.7$. The value of $\omega$ was found during the least-squares fit and then was fixed in order to reduce possible correlations between parameters. The use of the two-gap model in comparison to the single-gap model slightly reduces the value of $\chi^{2}$ with almost the same number of degrees of freedom. In the case of Mo$_{8}$Ga$_{41}$, the two-gap model implies the existence of two \textit{s}-wave superconducting gaps with similar energies and almost equal contributions ($\omega=0.7$). Certainly, this model replicates the single-gap BCS-type model when describing the $\lambda^{-2}(T)$ data, and the fitting results are very similar (compare the solid and dashed lines in Fig.~\ref{f7}). However, keeping in mind the non-BCS-type features listed above, we exclude the single-gap scenario from the consideration. At the same time, the $\lambda^{-2}(T)$ data corroborate the multigap behavior of Mo$_{8}$Ga$_{41}$ and reveal that the introduction of two superconducting gaps is sufficient to explain superconducting behavior of this compound. Note that similar behavior was reported for SrPt$_{3}$P, which was suggested to be a two-band superconductor with equal gaps\cite{kh2014}. Future studies that will help in resolving the gap structure in Mo$_{8}$Ga$_{41}$ are highly desirable.

\subsection{V for Mo substitution in Mo$_{7}$VGa$_{41}$}

In our previous study\cite{our}, Mo$_{\text{8-x}}$V$_{\text{x}}$Ga$_{41}$ was also found to be superconducting with the transition temperature $T_{\text{c}}$ being just slightly lower than that for the parent compound. Zero-field and transverse-field $\mu$SR experiments were conducted therefore on Mo$_7$VGa$_{41}$ having $T_{\text{c}}\simeq9.2$\,K at zero magnetic field\cite{our}. The analysis of the experimental data was performed similarly to that described previously for Mo$_{8}$Ga$_{41}$ sample.

\begin{figure*}
\includegraphics{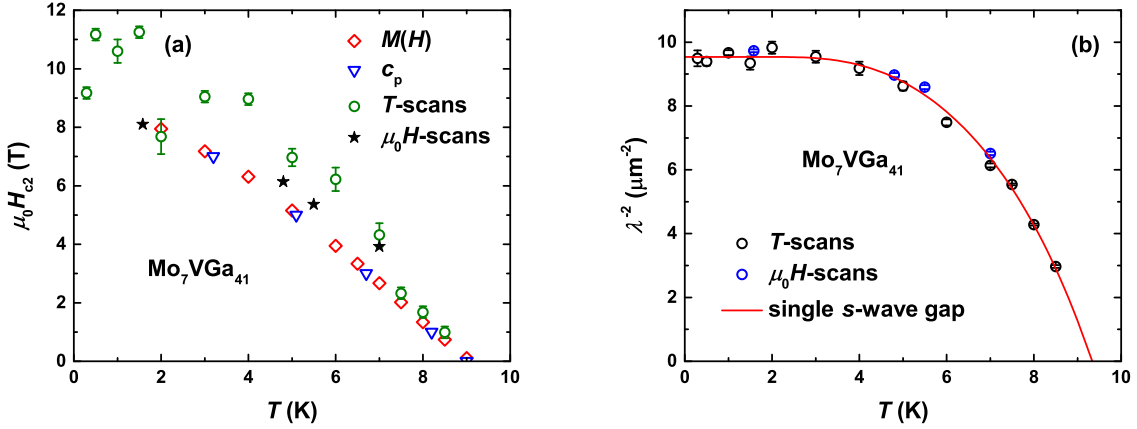}
\caption{\label{f8}(a) Upper crititcal field and (b) magnetic penetration depth of Mo$_{7}$VGa$_{41}$ plotted as $\lambda^{-2}$ \textit{vs.} $T$. Solid red line is a fit of the data according to the single-gap BCS-type model.}
\end{figure*}

Figure~\ref{f8}(a) shows a summary of the temperature-dependent upper critical field of Mo$_{7}$VGa$_{41}$. The $\mu_{0}H_{\text{c2}}$ data from the magnetization and heat capacity experiments as well as that obtained in TF-$\mu$SR studies are presented. In contrast to the unsubstituted Mo$_{8}$Ga$_{41}$, the TF-$\mu$SR $\mu_{0}H_{\text{c2}}$ values agree well with the data from thermodynamic measurements for Mo$_{7}$VGa$_{41}$. This implies that only one characteristic length scale is needed to describe the upper critical field. In accordance with this observation, the temperature dependence of the magnetic penetration depth of Mo$_{7}$VGa$_{41}$ presented as the $\lambda^{-2}(T)$ plot in Fig.~\ref{f8}(b) was satisfactorily fitted within the single-gap BCS-type model using Eq.~\eqref{eq4}. The fit yields $\Delta(0)=1.87(9)$\,meV and $\lambda(0)=292(4)$\,nm with the value of $\Delta(0)/k_{B}T_{\text{c}}=2.36$ exceeding the weak-coupling BCS limit of 1.76. Thus, from the TF-$\mu$SR data we suggest a simple \textit{s}-wave superconducting state in Mo$_{7}$VGa$_{41}$ accompanied by the strong electron-phonon coupling.

\subsection{Comparison of Mo$_8$Ga$_{41}$ with its V-substituted analog Mo$_7$VGa$_{41}$}

\begin{figure}
\includegraphics{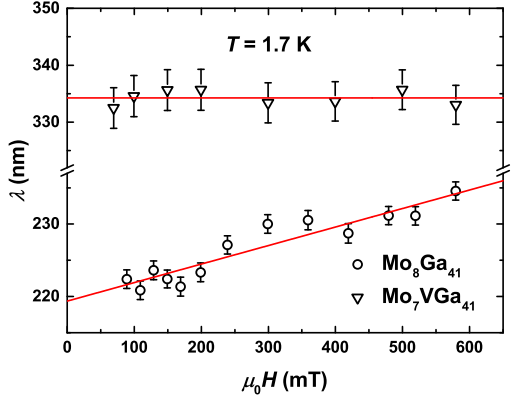}
\caption{\label{f9}Magnetic penetration depth $\lambda$ as a function of magnetic field for Mo$_{8}$Ga$_{41}$ (open circles) and Mo$_{7}$VGa$_{41}$ (open triangles). The solid red lines are linear fits of the data.}
\end{figure}

The experimental data presented above reveal that Mo$_{8}$Ga$_{41}$ and Mo$_{7}$VGa$_{41}$ are different in two important aspects. First of all, application of the equation of Brandt leads to different results. The Brandt equation (Eq.~\ref{eq3}) is formulated for the case of single-gap \textit{s}-wave superconductivity. By this reason, the values of $\mu_{0}H_{\text{c2}}$ extracted for Mo$_{7}$VGa$_{41}$ using Eq.~\ref{eq3} are in good agreement with thermodynamic measurements, since Mo$_{7}$VGa$_{41}$ exhibits actually a single-gap \textit{s}-wave superconductivity. The different situation is observed for Mo$_{8}$Ga$_{41}$. In this case, Eq.~\ref{eq3} yields the values of $\mu_{0}H_{\text{c2}}$, which are significantly lower than those from thermodynamic measurements, indicating that Mo$_{8}$Ga$_{41}$ is not a single-gap superconductor. Fourier analysis of the asymmetry spectra reveals no signs of disorder of the vortex lattice that may cause the Brandt equation being not applicable. The $\lambda(\mu_{0}H)$ dependence calculated using Eq.~\ref{eq3} also reveals the different behavior for Mo$_{8}$Ga$_{41}$ and Mo$_{7}$VGa$_{41}$ (Fig.~\ref{f9}). The field dependence of $\lambda$ observed for Mo$_{8}$Ga$_{41}$ is reminiscent of that of NbSe$_{2}$\cite{nbse2-1}, which is also a two-gap superconductor\cite{nbse2-2}. At the same time, there is no field dependence in the case of Mo$_{7}$VGa$_{41}$ in compliance with its conventional behavior. The substitution of just one out of eight Mo atoms by one V atom leads to the \textit{complete} suppression of one of the superconducting energy gaps.

The second aspect is that the V substitution is accompanied by more than a factor of 2 reduction of the superfluid density $\rho_{\text{s}}\propto\lambda^{-2}$. Indeed, from the fits of the experimental data we get $\lambda^{-2}(0)\simeq 21$ and 12\,$\mu\text{m}^{-2}$ for Mo$_{8}$Ga$_{41}$ and Mo$_{7}$VGa$_{41}$, respectively.

In order to explain such discrepancies, three different scenarios are going to be compared. According to the first one, the structural disorder caused by mixing Mo and V atoms in one crystallographic position may lead to the reduction in the superfluid density. From the single-crystal x-ray diffraction data\cite{our}, it is known that the substitution occurs in two crystallographic positions, and the average V content is less than 15\,at.\,\%. Such small amount of disorder in the crystal structure should not cause a drastic change of $\rho_{\text{s}}$. Moreover, as in the case of the K$_{\text{1-x}}$Na$_{\text{x}}$Fe$_{2}$As$_{2}$ solid solution \cite{knafese}, the reduction in $\rho_{\text{s}}$ caused by structural disorder should be accompanied by the significant reduction in $T_{\text{c}}$, whereas the Mo$_{\text{8-x}}$V$_{\text{x}}$Ga$_{41}$ solid solution demonstrates no significant change of $T_{\text{c}}$. By substituting V for Mo, the transition temperature just slightly reduces from $T_{\text{c}}=9.8$\,K in Mo$_{8}$Ga$_{41}$ to 9.2\,K in Mo$_{7}$VGa$_{41}$\cite{our}.

In contrast to K$_{\text{1-x}}$Na$_{\text{x}}$Fe$_{2}$As$_{2}$, the V for Mo substitution is not isovalent. Rather, the change in the electron count is observed, which enables us to formulate the second scenario, with the heterovalent substitution being responsible for the reduction in $\rho_{\text{s}}$. The effect of heterovalent substitution on the electronic structure of Mo$_{\text{8-x}}$V$_{\text{x}}$Ga$_{41}$ was analyzed in our previous study\cite{our}. By using the rigid-band shift approximation, it was shown that no significant change in the density of states at the Fermi level occurs with increasing V content in the solid solution. Thus, the heterovalent nature of the substitution also does not explain the observed reduction in $\rho_{\text{s}}$.

Finally, the third scenario considers competing states that cause the reduction in the superfluid density $\rho_{\text{s}}$. The Ba(Fe$_{\text{1-x}}$Co$_{\text{x}}$)$_{2}$As$_{2}$ solid solution\cite{bafecoas} is an example, in which the competition between superconductivity and magnetism is observed. On the phase diagram of Ba(Fe$_{\text{1-x}}$Co$_{\text{x}}$)$_{2}$As$_{2}$, the superconducting dome is intersected with the spin-density-wave field yielding the intermediate region of competing states. Measurements of the magnetic penetration depth $\lambda$ clearly show that entering the intersected region results in the drastic increase in $\lambda$, which is equivalent to the reduction in $\rho_{\text{s}}$\cite{bafecoas}.

\begin{figure}
\includegraphics{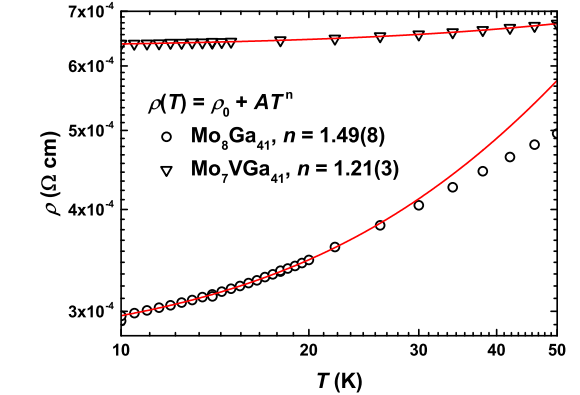}
\caption{\label{f10}Resistivity of Mo$_{8}$Ga$_{41}$ (open circles) and Mo$_{7}$VGa$_{41}$ (open triangles) at low temperatures. The data are taken from ref.~[8]. Solid red lines are the results of fitting (see in the text).}
\end{figure}

At this point, we make an assumption that the emergence of some competing state in Mo$_{7}$VGa$_{41}$ may lead to the suppression of one of the superconducting gaps accompanied by the significant reduction in the superfluid density. The collected ZF-$\mu$SR data show that both Mo$_{8}$Ga$_{41}$ and Mo$_{7}$VGa$_{41}$\cite{sm} lack spontaneous coherent magnetic fields in the superconducting state, thus excluding magnetic order as a possible competing state from the consideration. To obtain further information, we analyze the low-temperature resistivity data\cite{our} for Mo$_{8}$Ga$_{41}$ and Mo$_{7}$VGa$_{41}$. The $\rho(T)$ data at low temperatures can be fitted using equation $\rho(T)=\rho_{0}+AT^{n}$ (Figure~\ref{f10}). The fitting yields the exponents $n=1.49(8)$ for Mo$_{8}$Ga$_{41}$ and $n=1.21(3)$ for Mo$_{7}$VGa$_{41}$. This reduction in the exponent upon increasing V content may indicate the onset of an incipient charge-density-wave state in Mo$_{7}$VGa$_{41}$. A similar situation has been observed in Ta$_{4}$Pd$_{3}$Te$_{16}$\cite{tapdte}, where the suppression of $n$ under pressure signifies enhancement of CDW fluctuations. The presence of CDW state in the V-substituted Mo$_{8}$Ga$_{41}$ would make one band unavailable for superconductivity and reduce the density of the supercarriers due to the opening of the energy gap associated with the CDW regime.

\section{Conclusions}

A comprehensive study by means of muon spin rotation/relaxation technique was carried out on a strongly-coupled superconductor Mo$_{8}$Ga$_{41}$ and its V-substituted analog Mo$_{7}$VGa$_{41}$. ZF-$\mu$SR experiments show that both compounds lack any spontaneous coherent magnetic fields in the superconducting state. TF-$\mu$SR experiments elucidate microscopic properties of the superconducting state, which turns out to be substantially different in Mo$_{8}$Ga$_{41}$ and Mo$_{7}$VGa$_{41}$. The upper critical field of Mo$_{8}$Ga$_{41}$ extracted from the TF-$\mu$SR data is lower than in thermodynamic measurements indicating the multigap superconductivity that leads to two or more independent length scales in the superconducting state. Accordingly, the temperature-dependent magnetic penetration depth was approximated by the model, in which two \textit{s}-wave superconducting gaps with similar energies and almost equal contributions are assumed. The V for Mo substitution does not affect $T_{\text{c}}$ significantly, but leads to the complete suppression of one superconducting gap. This follows from the substantial reduction in the superfluid density $\rho_{\text{s}}$ and from the single-gap behavior of both the upper critical field and magnetic penetration depth of Mo$_{7}$VGa$_{41}$. We speculate that the emergence of a competing state in Mo$_{7}$VGa$_{41}$ is responsible for the closure of one of the superconducting gaps.

\begin{acknowledgements}
This work was performed at the Swiss Muon Source, Paul Scherrer Institute, Villigen, Switzerland. Part of the experimental work was done by Gustavo Prack, Judith Suter and Clemens Spinnler (Swiss Nanoscience Institute, University of Basel, Basel, Switzerland) within their lab course at PSI. R.K. acknowledges Stefan Holenstein and Jean-Christophe Orain for their help during the $\mu$SR experiments. The work has been supported by the Russian Science Foundation, grant \#17-13-01033. V.Yu.V. appreciates the support from the European Regional Development Fund, project TK134. A.A.T. is grateful for the financial support by the Federal Ministry for Education and Research under the Sofja Kovalevskaya Award of the Alexander von Humboldt Foundation.
\end{acknowledgements}

\bibliography{fulltext}
\end{document}